\documentclass[aps,twocolumn,epsf,floats,pre]{revtex4-1}
\usepackage{graphics,graphicx,epsfig}
\usepackage{amssymb,color}
\usepackage{epsf,epstopdf,wrapfig}
\usepackage{amsmath}
\usepackage{multirow}

\newcommand{\be}{\begin{equation}}

\newcommand{\ee}{\end{equation}}
\newcommand{\bea}{\begin{eqnarray}}
\newcommand{\eea}{\end{eqnarray}}
\newcommand{\norm}[1]{\left\lVert#1\right\rVert}

\begin{document}

\title{ Non-symmetric interactions trigger collective swings in globally ordered systems}

\author{ Andrea Cavagna$^{a,b}$, Irene Giardina$^{a,b}$, Asja Jelic$^{a,b,c}$, Edmondo Silvestri$^{a,d}$, Massimiliano Viale$^{a,b}$}

\affiliation{$^a$ Istituto Sistemi Complessi, Consiglio Nazionale delle Ricerche, UOS Sapienza, 00185 Rome, Italy}
\affiliation{$^b$ Dipartimento di Fisica, Universit\`a\ Sapienza, 00185 Rome, Italy}
\affiliation{$^c$ The Abdus Salam International Centre for Theoretical Physics, Strada Costiera 11, 34014 Trieste, Italy}
\affiliation{$^d$ Dipartimento di Fisica, Universit\`a\ di Roma 3, 00146 Rome, Italy}

\begin{abstract}
Many systems in nature, from ferromagnets to flocks of birds, exhibit ordering phenomena on the large scale. In physical systems order is statistically robust for large enough dimensions, with relative fluctuations due to noise vanishing with system size. Several biological systems, however, are less stable than their physical analogues and spontaneously change their  global state on relatively short timescales.  In this paper we show that there are two crucial ingredients in these systems that enhance the effect of noise, leading to collective changes of state: the non-symmetric nature of interactions between individuals, and the presence of local heterogeneities in the topology of the network. The consequences of these features can be larger the larger the system size leading to a localization of the fluctuation modes and a relaxation time that remains finite in the thermodynamic limit. 
The system keeps changing its global state in time, being constantly driven out of equilibrium by spontaneous fluctuations. Our results explain what is observed in several living and social systems and are consistent with recent  experimental data on bird flocks and other animal groups.

\end{abstract}

\maketitle

Ordering phenomena are ubiquitous in nature, spanning from ferromagnetism and structural transitions in condensed matter, to collective motion in biological systems, and consensus dynamics in social networks. Order by itself requires a notion of robustness: the degree of global coordination must be stable in spite of noise, at least on certain time scales. This concept is quantified rigorously in equilibrium statistical physics: a system exhibits long range order when the relative fluctuations of the global order parameter (the observable quantifying order) are vanishingly small in the thermodynamic limit.  In a finite system, due to noise,  global order (e.g. the magnetization in a ferromagnet) can fluctuate, but such fluctuations are so small that bringing the system away from its original state would take a huge amount of time, the longer the larger the size of the system.

Many biological systems displaying ordered patterns, however, exhibit a larger sensitivity to noise than their physical analogues, and can change their  state on relatively short timescales.  Flocks of birds, for example,  have very large polarization but they turn and change spontaneously their flight direction very frequently \cite{attanasi+al_14,attanasi+al_15}. 
Consensus in social networks can swiftly  switch from a selected choice to another \cite{centola_10}.  Fluctuations appear to have a dominant role and one might wonder what kind of mechanism is responsible for this behavior, and whether it implies a disruption of long range order in the statistical physics sense.

In this paper we investigate this problem and show that there are two crucial ingredients that enhance the effect of noise leading to collective changes of state: the non-symmetric nature of interactions between individuals, and the presence of local heterogeneities in the topology of the interaction network. 
Surprisingly, the consequences of these two features are not limited to finite systems. 
Rather, their effect can be larger the larger the system size leading to a 
localization of the fluctuation modes and a relaxation time that remains finite in the thermodynamic limit. The system keeps changing its global state in time, being constantly driven out of equilibrium by spontaneous fluctuations. 
Non-symmetric interactions and  network heterogeneities naturally occur in many biological instances of collective behavior, where the individual units coordinate with each other in a non reciprocal way and the local connectivity is different at the boundary and in the bulk. Our analysis therefore explains why such systems exhibit the sensitivity to fluctuations observed in experiments. Besides, we show that big fluctuations typically build up and start at the boundary, peripheral nodes acting as triggers for the global change.

Let us start by considering the archetypical case of global order in a physical system: the ferromagnet. The minimal description of a ferromagnet is given  by the standard Heisenberg model where vectorial spins are placed on a $d$-dimensional lattice and interact via nearest-neighbors alignment interactions. The Hamiltonian of the system reads:
\be
{\cal H}=- J \sum_{ij} n_{ij} {\vec \sigma}_i \cdot {\vec \sigma}_j \ \ \ \ \ \ \norm{\vec\sigma}=1.
\label{Heisenberg}
\ee
Here the $\{{\vec \sigma}_i\}$ are norm one continuous vectors,  and $n_{ij}$  - the adjacency matrix - is equal to $1$ for interacting spins (i.e. 
neighboring sites in the lattice) and $0$ otherwise. 
The Heisenberg model also offers the simplest case of imitative interaction rules, which are commonly used to model biological and social groups \cite{couzin+krause_03,vicsek_review,bialek+al_12}. 

For $d>2$, the Heisenberg model has an ordering transition at finite temperature $T_c$. For $T<T_c$  the system exhibits a spontaneous magnetization $\vec{M}=(1/N)\sum_i {\vec{\sigma_i}}>0$, all spins pointing on average in the same direction. For a finite system of size $N$, the relaxation time $\tau_N$ -  the time needed for the system to change the direction of the magnetization - grows with the system size, ensuring stability of order in the thermodynamic limit. More specifically \cite{goldschmidt_86,niel_86},
\be
\tau_N \sim N
\label{tau_sym} \ .
\ee

In the low temperature regime a simple computation  illustrates well the mechanism of relaxation and the scaling with size. 
Let us consider the case of planar spins on a $d=3$ lattice, which is simpler to handle algebraically  (the computation is easily generalized). Each spin can be described by a phase $\varphi_i$, measuring its angle with respect to the global magnetization $\vec{M}$. To compute the relaxation time, we need to specify what is the dynamics followed by the system.  To keep the problem as general as possible we consider both dissipative and inertial dynamical terms, covering a variety of dynamical behaviors in physical \cite{HH_77,goldschmidt_86,niel_86,chaikin_00} and biological \cite{vicsek_95,vicsek_review,attanasi+al_14,attanasi+al_15} ordered systems. 
At low temperature $M\sim 1$, the phases are small, and the dynamical equations read (see SM)
\be
\chi \frac{d^2 \varphi_i}{d t^2}=-J\sum_j \Lambda_{ij}\varphi_j-\eta\frac{d\varphi_i}{dt}+\xi_i
\label{eq:dyn}
\ee
where $\Lambda_{ij}=-n_{ij}+\delta_{ij} \sum_{k} n_{ik}$ is the discrete Laplacian. 
$\chi$ and  $\eta$ represent, respectively, a rotational inertia and a rotational viscosity; and $\xi_i$ is a random Gaussian noise  
with $\langle \xi_i(t) \xi_j(t^\prime) \rangle = \Delta  \eta \ \delta_{ij}\delta(t-t^\prime)$, $T=\Delta/2$ being the temperature of the asymptotic equilibrium distribution. 
By taking the limit $\chi/\eta^2 \to 0$  we recover a purely overdamped dynamics (as in \cite{goldschmidt_86,niel_86,vicsek_95}), while $\eta\to 0$ corresponds to a reversible Hamiltonian dynamics (see e.g. \cite{chaikin_00,attanasi+al_14,attanasi+al_15}).

Let us assume that at time $t=0$ the system is in a strongly polarized state with $\vec{M}=\vec{M}_0$. The magnetization then fluctuates in time due to the spontaneous noise acting on the system. To evaluate the relaxation time, we compute the perpendicular fluctuation $\delta {\vec M}^{\perp}$ of the magnetization with respect to $\vec{M}_0$, measuring how much the state of the system has departed from the original direction. One finds (see SM)
\be
\langle \left (\delta M^\perp \right )^2\rangle = D_0 \frac{t}{\eta} 
\label{perp}
\ee
where averages are taken over the dynamical noise, and the diffusion coefficient $D_0$ is given by
\be
D_0= \frac{\Delta}{N} \sum_i  (w^0_i)^2
\label{dsym}
\ee 
Here ${\bf w^{0}}$ is the $N$-dimensional lowest eigenvector of the Laplacian matrix. This eigenvector corresponds to a zero mode, the so called Goldstone mode, resulting from the rotational symmetry of the Hamiltonian. It can be easily shown that ${\bf w^0}$ is a constant vector, with $w^0_i=1/\sqrt{N}$, giving a diffusion coefficient $D_0\sim 1/N$. When the perpendicular component (\ref{perp}) becomes of order $1$ the system has changed its global direction, i.e. it has relaxed from the original state. This occurs when $t/\eta \sim 1/D_0\sim N$, giving the scaling  of the relaxation time with size of Eq.~(\ref{tau_sym}). 

The interpretation of this result is straightforward but illuminating. When the system orders, it spontaneously choses a direction among all the possible ones. There remain however `easy fluctuations' in the manifold perpendicular to $\vec{M}$ that is described by the zero eigenspace of the Laplacian.  When the system evolves in presence of noise, it moves along these soft modes: fluctuations are small but build up in time leading to the diffusive behavior of the magnetization. Due to the homogeneity of the interaction network fluctuation modes are delocalized: each spin equally contributes to the global fluctuations with a vanishing weight leading to the divergence of the relaxation time with the system size, and to stability of order in the thermodynamic limit. This behavior is  ensured by the symmetry of the interaction matrix $n_{ij}$.

\begin{figure*}[t]
\includegraphics[width=0.65\columnwidth]{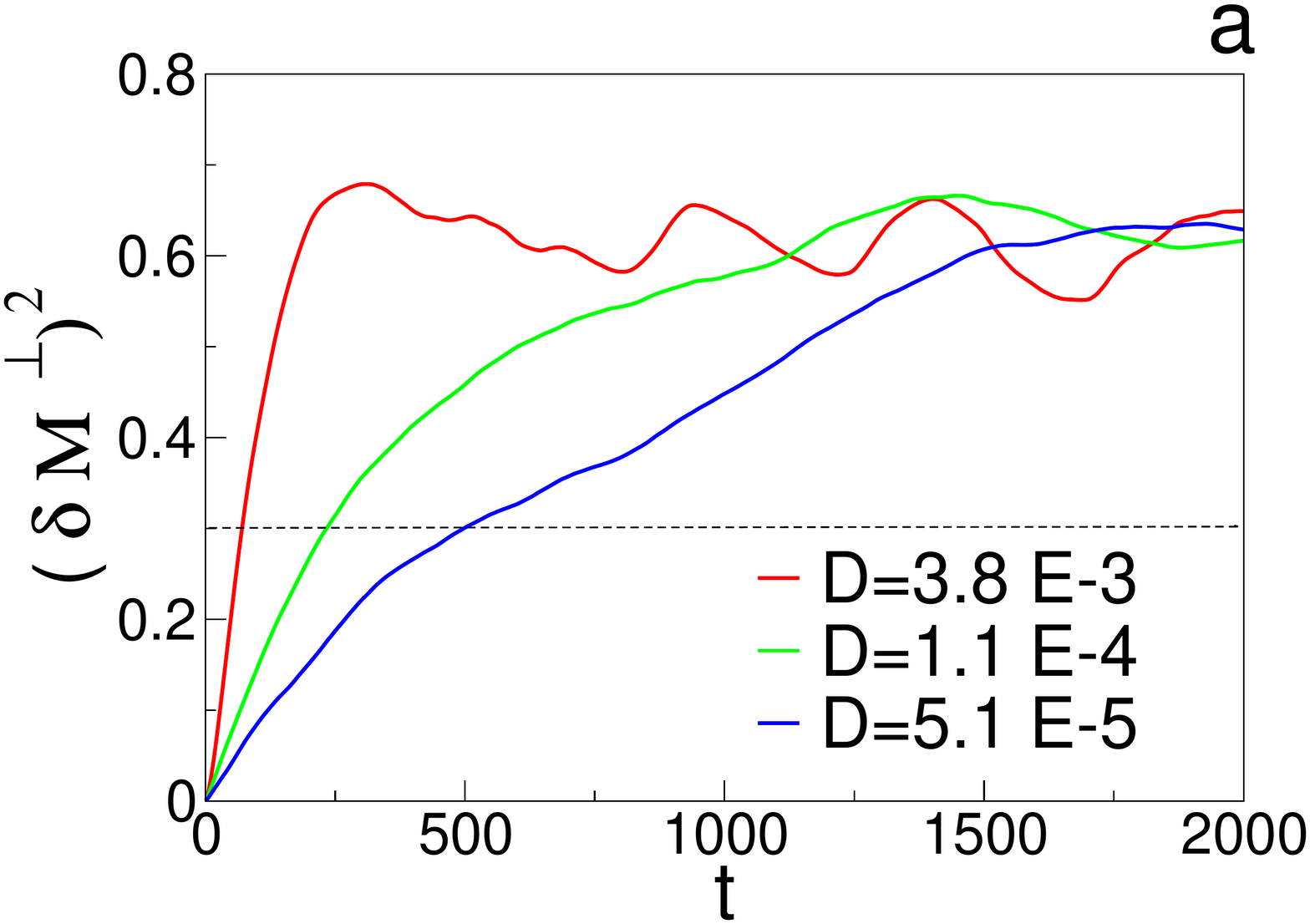}\ \ \
\includegraphics[width=0.65\columnwidth]{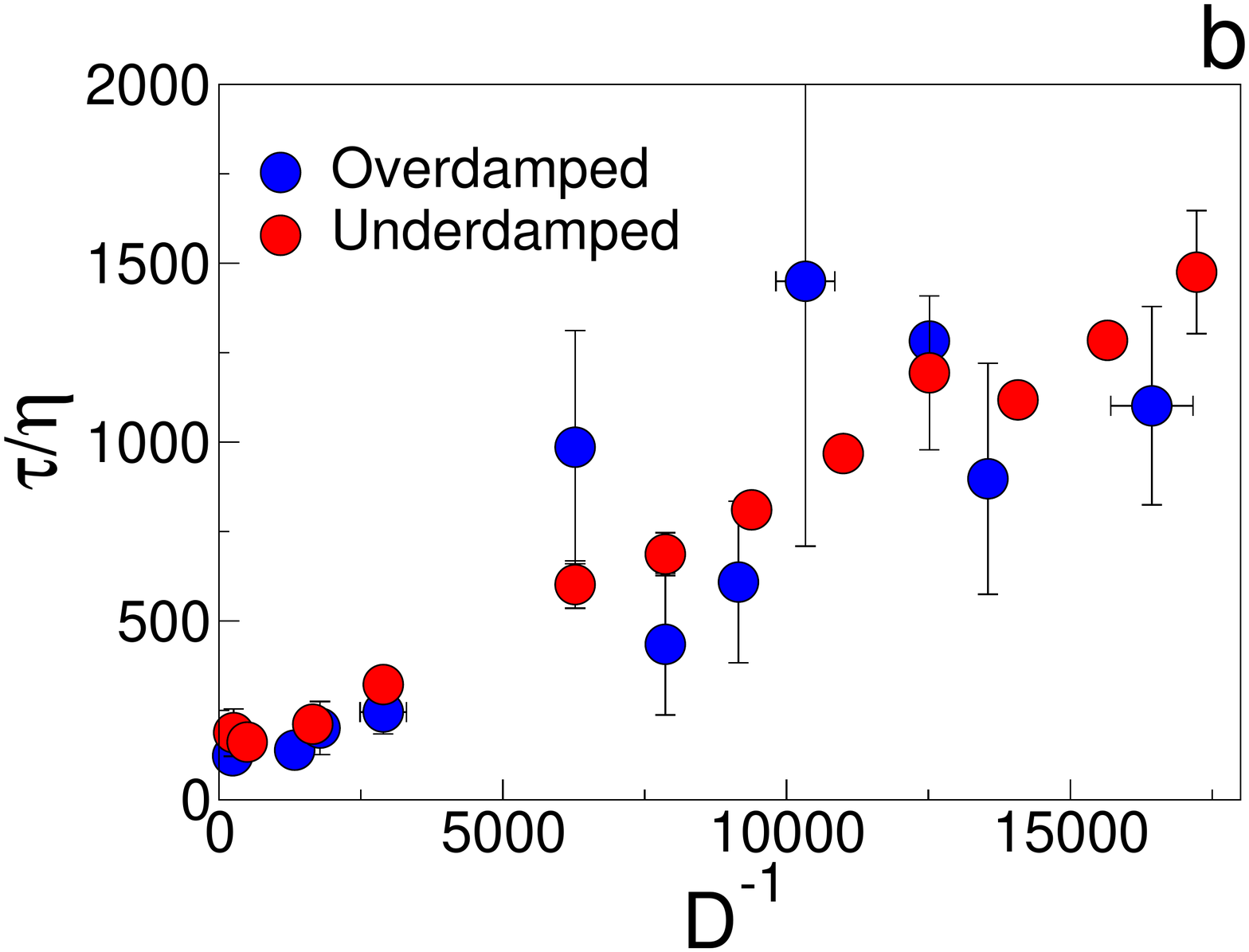}\ \ \
\includegraphics[width=0.65\columnwidth]{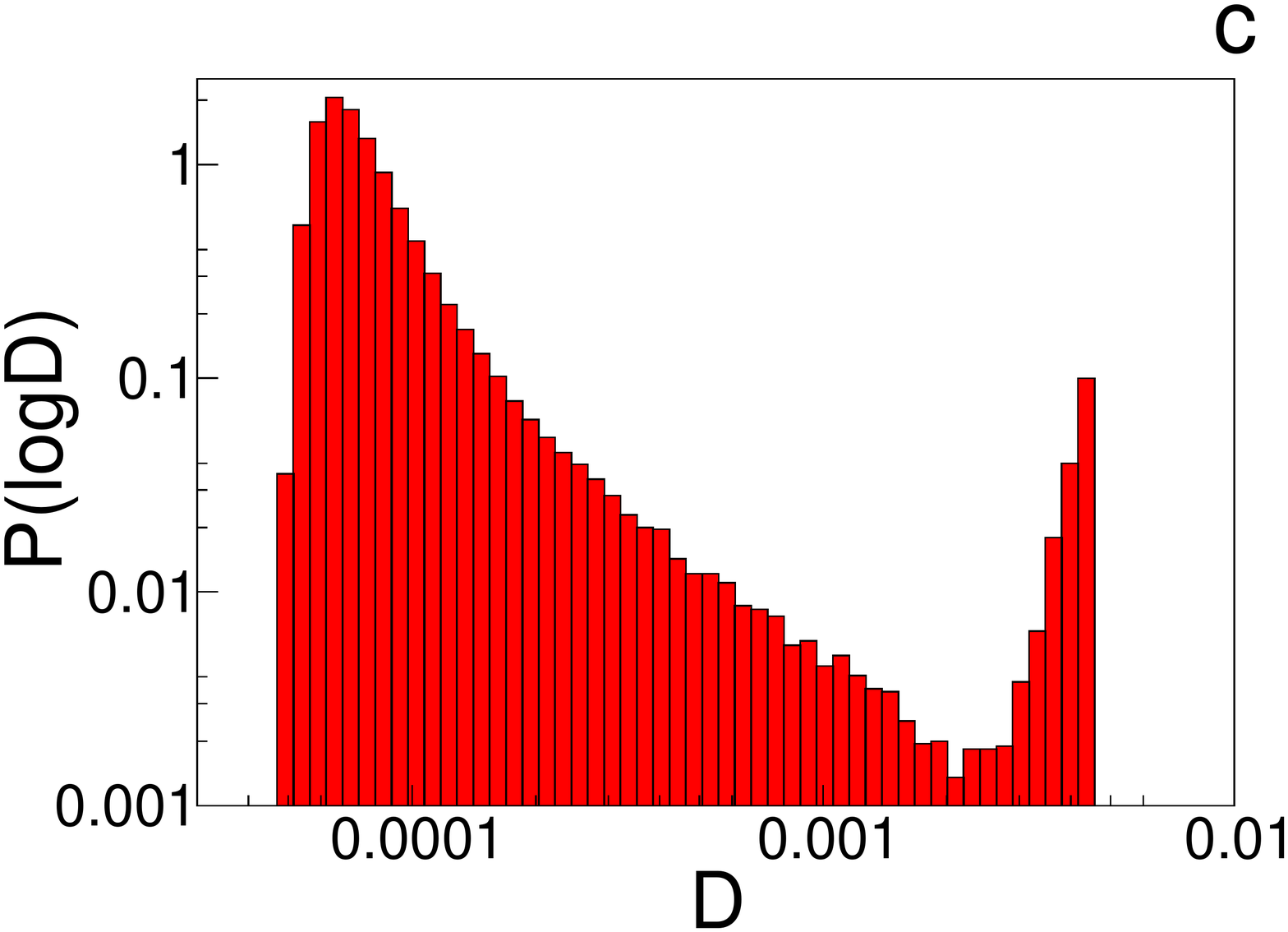}
\caption{a) Perpendicular fluctuation of the global magnetization as a function of time, for three networks with different diffusion coefficients in the NERH ensemble with $N=1000$ and $n_c=6$ (underdamped dynamics). The relaxation time is defined as the time where   $\langle \delta M^\perp(t)^2\rangle=0.3$ (black dotted line). b) Relaxation time vs diffusion coefficient, for the same ensemble, with underdamped and overdamped dynamics.
c) Distribution Probability of the diffusion coefficient $D$ for the $N=1000$, $n_c=6$ ensemble; we plot the distribution of $\rm{log}(D)$ to better visualize the secondary peak. The parameters of the dynamics are: $J=1.2$, $\Delta=0.032$, $\chi=0.83$, $\eta=15$ (overdamped dynamics), $\eta=0.3$ (underdamped dynamics).
 }
\label{fig:probtau}
\end{figure*}

Let us now move beyond the standard paradigm described above, by introducing in a minimal way a few crucial features characterizing real biological and social networks. First, biological and social interactions are not necessarily symmetric. Animals in a group, for example, usually perceive or receive signals from neighbors who are not themselves able to see them. Second, the lattice structure is rather restrictive since many biological systems do not exhibit any kind of structural order in space \cite{couzin+krause_03,cavagna+al_08}.  
The simplest thing we can do is to draw points uniformly in Euclidean space instead of using the sites of a lattice.  Then we can place the spins on each of such points and prescribe (as in the original Heisenberg model) that a spin interacts with its first $n_c$ neighbors in space. 
Since the neighborhood relationship is in general non-reciprocal, the resulting interactions are automatically non symmetric.  
What we get is an alignment model on a random Euclidean network,  with a non-symmetric adjacency matrix $n_{ij}$ (where $n_{ij}=1$ if $j$ is one of the first $n_c$ nearest neighbors of $i$ and $0$ otherwise). For $d=3$ and $n_c=6$ this system has the same dimensionality and connectivity as the standard Heisenberg model: each spin interacts exactly with $n_c$ neighbors. However, since $n_{ij}\ne n_{ji}$ there can be local unbalances in the interaction network where an individual spin is `seen' by fewer or more spins than those it interacts with, i.e. $\sum_j n_{ij}=n_c\ne \sum_j n_{ji}$.  Interestingly, natural flocks of birds exhibit a similar typology of alignment interaction network  \cite{ballerini+al_08,bialek+al_12,cavagna+al_15,mora+al_15} and their behavior is described by a dynamics of the kind of (\ref{eq:dyn}) in the underdamped limit \cite{attanasi+al_14,cavagna+al_14,cavagna+al_15b}.
We note that interactions being non-symmetric, detailed balance is not obeyed and we can expect off-equilibrium features \cite{zwanzig,naomi}.
  Besides, the adjacency matrix $n_{ij}$ is now a random matrix belonging to the class of non-Hermitian Euclidean Random Matrices \cite{mezard,skipetrov_11}, with non trivial spectral properties. The model we have defined - that we call the Non-Symmetric Euclidean Random  Heisenberg model  (NERH) - might therefore lead to novel dynamical behavior.

Let us now investigate whether and how the relaxation properties of the system change, due to the non-symmetric nature of the interaction network. The  computation of the relaxation time leading to Eqs. (\ref{perp}) and (\ref{tau_sym}) can be extended to this more general case. Both the adjacency matrix $n_{ij}$ and the discrete Laplacian $\Lambda_{ij}$ are now non-symmetric. 
As a consequence, they have right and left eigenvectors, which behave differently. The analogous of Eqs. (\ref{perp})(\ref{dsym}) are 
\be
\langle \left (\delta M^\perp \right )^2 \rangle = D \frac{t}{\eta} + F_{dyn}(t)
\label{perp2}
\ee
where $F_{dyn}(t)$ is a sub-dominant (in time) contribution, which depends on the specific dynamics considered (see SM). The diffusion coefficient is given by
\be
D = \frac{\Delta}{N} \sum_i (u^0_i)^2  \ .
\label{dasym}
\ee 
where $\bf u^0$ is the left zero eigenmode. Contrary to the symmetric case, $\bf u^0$ is not a constant vector and depends on the specific network considered. 
Since the term in $D$ dominates the evolution of the global fluctuations in Eq.(\ref{perp2}), also in this case we have 
\be
\frac{\tau}{\eta} \sim \frac{1}{D}
\label{relax}
\ee

\begin{figure*}[t]
\includegraphics[width=0.65\columnwidth]{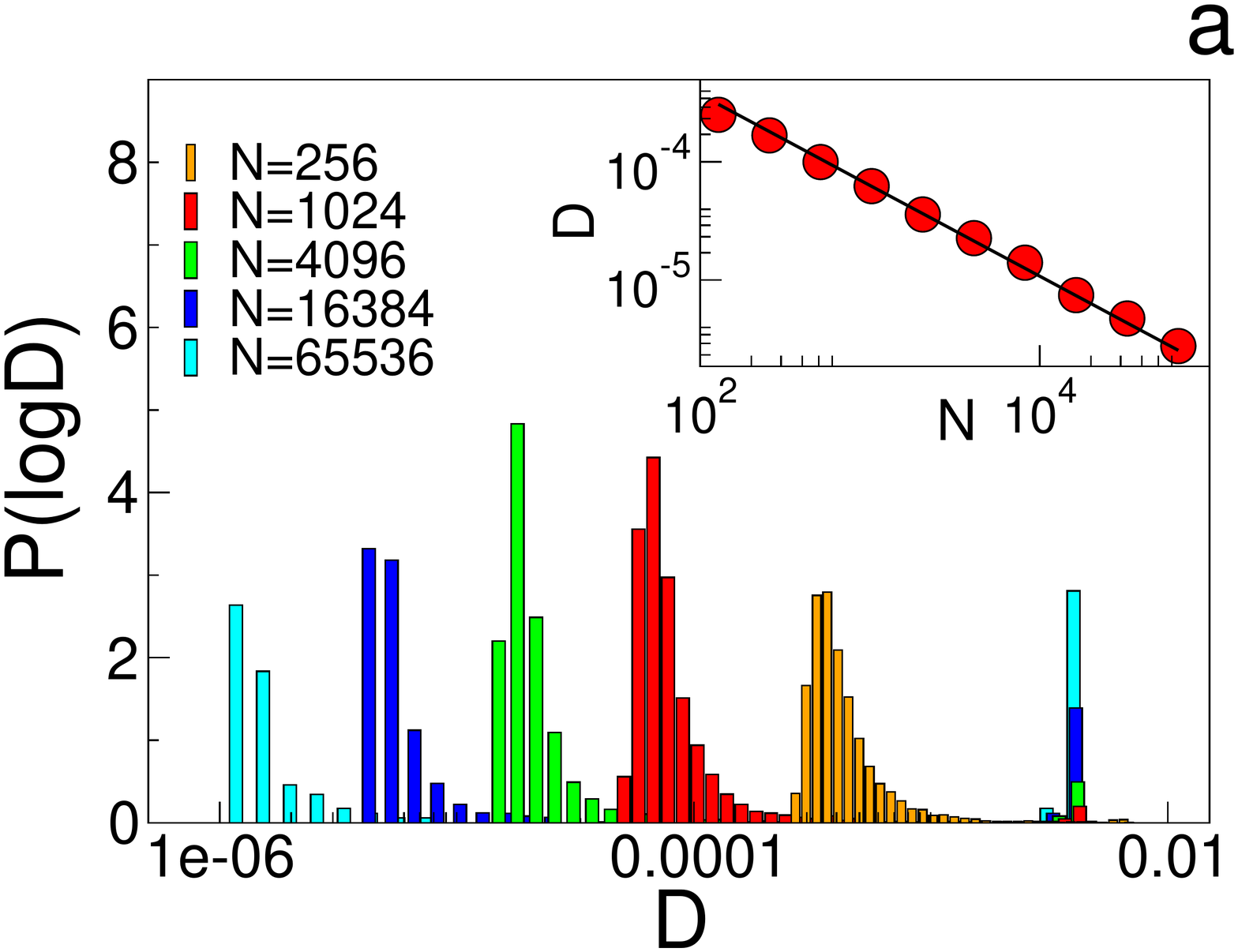}
\includegraphics[width=0.65\columnwidth]{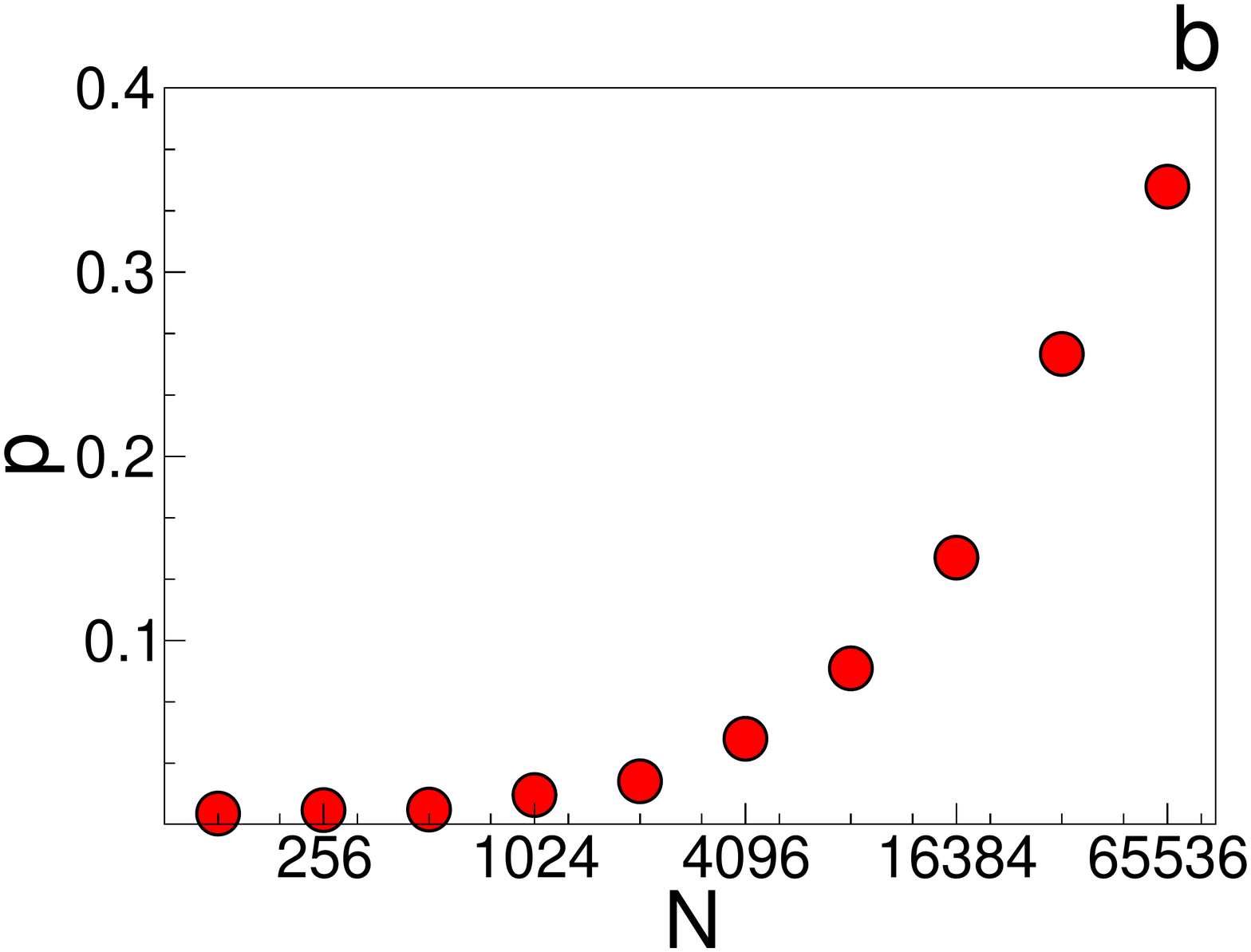}
\includegraphics[width=0.65\columnwidth]{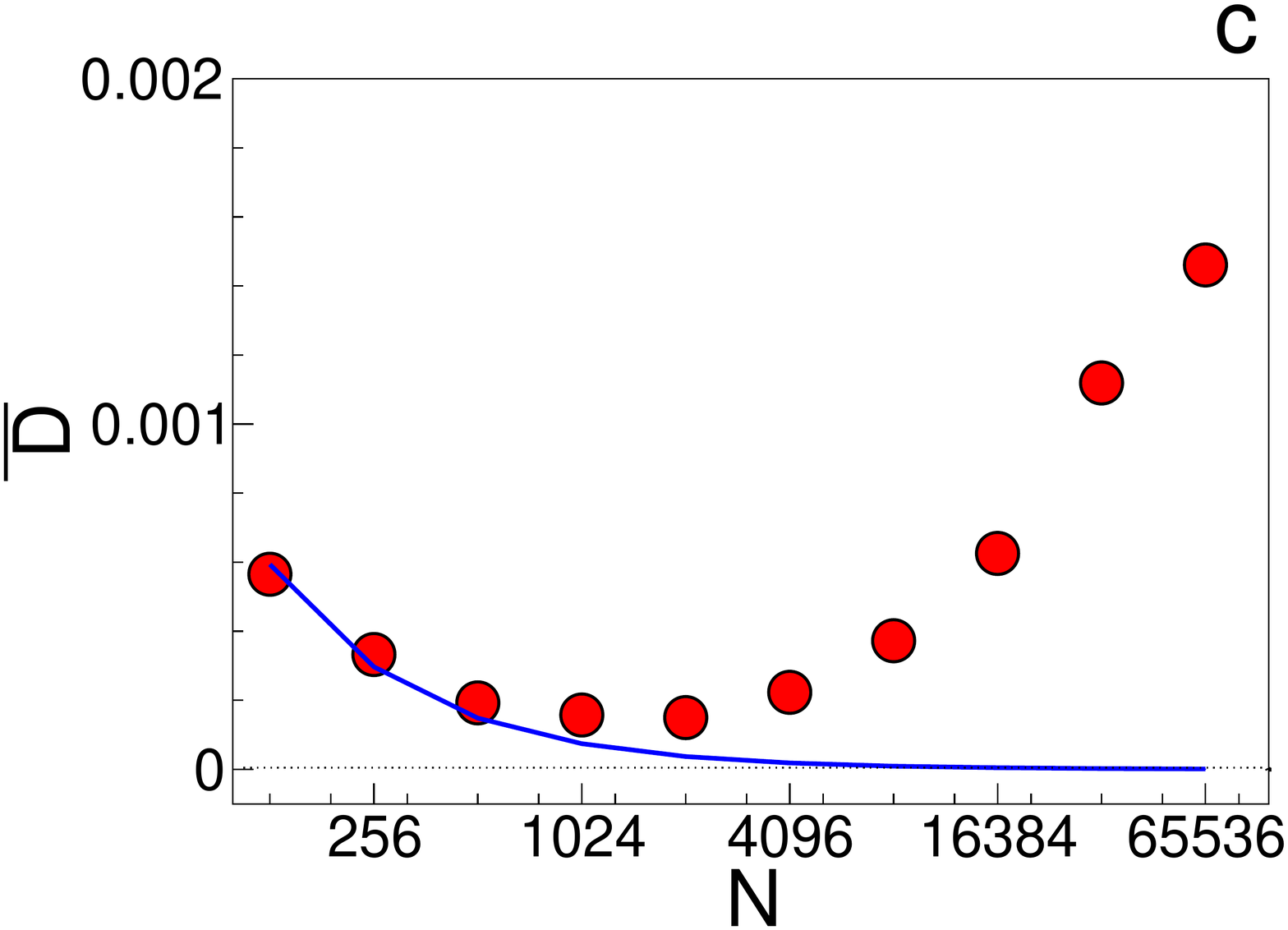}
\caption{a) Probability distribution of the diffusion coefficient for ensembles of NERH of  different size. Inset:  value of $D$ at the primary peak, as a function of the network size. b) Probability of the secondary peak as a function the network size. 
 c)  Average diffusion coefficient as a function of the network size; the blue line represents the $1/N$ behavior expected for homogeneous networks. $N\in [128: 65536]$; $n_c=6$. Qualitatively similar results also hold for different values of $n_c$ - see SM.}
\label{fig:scaling}
\end{figure*}

These equations once again connect the relaxation of the system (a dynamical quantity) with the spectral properties of the Laplacian (a topological feature of the network). Now,  however,  different networks have different $\bf u^0$, different $D$ and, consequently, different relaxation properties. To check the validity of Eqs. (\ref{perp2}-\ref{relax})
 and explore their consequences, we have generated many samples of NERH networks of size $N$. Each network was obtained by drawing at random $N$ points in a $3D$ volume, and building the  asymmetric spin-spin interaction graph as discussed above. For each network we computed the lowest zero left eigenmode $\bf u^0$, and evaluated the diffusion coefficient as defined in (\ref{dasym}). Then, given a network, we performed a numerical simulation of the dynamical evolution of the NERH (see Eq.~(\ref{eq:dyn}) and SM). We considered the system in the low temperature region ($M\sim 0.98$)  and evaluated the relaxation time $\tau$ as the time where $\langle (\delta M^\perp(t))^2\rangle\sim O(1)$ (see Fig.~\ref{fig:probtau}a).  In Fig.~\ref{fig:probtau}b we plot the relaxation time (computed from the dynamics) as a function of the diffusion coefficient (computed from the network) for networks with $N=1024$. This figure shows that  the relaxation time indeed scales inversely with the diffusion coefficient. Besides this behavior occurs independently of the dissipative/reversible character of the implemented dynamics, confirming that the second term in Eq.~(\ref{perp2}) is subdominant.

Equations (\ref{perp2}-\ref{relax}) 
show that the relaxation time of the system crucially depends on the properties of the left eigenvector $\bf u^0$. In particular, $u^0_i$ - also known as the {\it eigenvalue centrality} of node $i$ \cite{bonacich_01}- can vary from node to node determining different contributions to the global fluctuations and an overall different value of the diffusion coefficient.  If $\bf u^0$ is extended (similarly to what happens in a regular lattice) then centrality is homogeneously distributed through the network, $D\sim 1/N$ and the relaxation time is proportional to the size $N$ of the system. If, however, $\bf u^0$ is localized on a finite subset of nodes (i.e. some nodes are significantly more central than others) the diffusion coefficient could be substantially larger leading to  much shorter relaxation times.  The distribution  of $D$ in the network ensemble for $N=1024$ is plotted in Fig \ref{fig:probtau}c.
We can see that this distribution has a large main peak centered on the value $D\sim 1/N$ (the same we would get in a symmetric interaction network of the same size) indicating that most of the networks behave in a homogeneous manner and have a small diffusion coefficient, as in the Heisenberg model on a regular lattice. There are however a few networks  with a diffusion coefficient that is substantially larger,  corresponding to the secondary peak of the distribution. The homogenous networks with small $D$ have long relaxation times, while the few ones with large $D$ relax on much quicker scales (see Fig.~\ref{fig:probtau}a,b).

These results indicate that we have a bimodality in the distribution of the diffusion coefficient and, consequently, of the relaxation time.  To understand the  relevance of this result we need to understand how the distribution $P(D)$ changes with the system size.
To explore the finite size scaling of the relaxation behavior, we have generated ensembles of NERH networks for different values of $N$ ranging from $N=128$ to $N=65536$. For each ensemble of size $N$, we computed the distribution $P(D)$. The resulting curves are plotted in Fig.~\ref{fig:scaling}a. What we see is that i) the primary peak is centered on a value of $D$ that decreases with system size (see inset) as would happen for symmetric networks; ii) the secondary peak is instead always peaked on the same finite value $D\sim \Delta/n_c$ and its height increases with the size $N$ of the network.  We also computed the global probability of finding a network with finite $D$, defined as the integral over the secondary peak. As can be seen from Fig.~\ref{fig:scaling}b this probability  increases with $N$: the occurrence of networks with significant topological heterogeneities and a finite diffusion coefficient is  not, therefore, a finite size effect. On the contrary, these networks are statistically more relevant the larger the system size.  

Let us now discuss how this scaling of the probability distribution affects relaxation at ensemble level. Clearly the dynamical relaxation of a given network of size $N$ is determined by the specific diffusion coefficient of that network: there are quick networks that change global magnetization over a short relaxation time (large $D$) and slow more stable networks that require a much longer time to relax (small $D$). But what happens on average? To answer this question we need to be careful. If we compute  the average relaxation time over all networks of size $N$, this time is dominated by the slow networks up to very large values of $N$ (even though quick networks exist and become increasingly more probable). Rather,  what we need to do is to compute directly the average diffusion coefficient, which well captures the weight of the quick networks. (This subtle distinction between averaging relaxation time vs. diffusion coefficient is well-known in other cases where two distinct dynamical behaviors occur, as for example in the violation of the Stokes-Einstein relation in supercooled liquids \cite{cavagna_review}). In  Fig.~\ref{fig:scaling}c. we therefore plot  the average diffusion coefficient $\overline{D}$ (the bar indicates averages over the network ensemble), as a function of the system size $N$. After an initial decrease, $\overline{D}$ reaches a minimum and then increases asymptotically with the size of the system: in the thermodynamic limit the diffusion coefficient does not go to zero, as in the symmetric case, but tends to a non vanishing value. We therefore conclude that non symmetric interactions can have a dramatic impact on the relaxation of the system, which persists in the thermodynamic limit: if we draw at random a 
NERH network this network will have, with finite probability, a finite relaxation time, no matter how large the network is.

Let us now discuss the physical reasons why some networks have a localized eigenvector $\bf u^0$, and why this localization has such impressive consequences on the collective dynamics of the system. If we compute the partecipation ratio $PR=1/N \sum_i (u^0_i)^2$ for the quick networks with finite diffusion coefficient, we find  $PR \sim 1/n_c$. This means that there are approximately $n_c$ nodes that dominate the collective fluctuations of the network. Besides, these most influent nodes are closely located in space and very connected one to the other, as quantified by the high value of the clustering coefficient $c_i=1/[n_c(n_c-1)]\sum_{jk} n_{ij}n_{ik}n_{jk}$. For example, for $N=1000$ we find  
$<c_i>=0.90 \pm 0.11$, to be compared to the value $<c_i>=0.61 \pm 0.13$ of the slow networks 
(see also Fig. \ref{fig:examples}a).
Since  the number of interacting neighbors of each node is limited to $n_c$, this highly clustered region tends to be poorly connected with the rest of network: all the nodes in the cluster see other nodes in the same cluster, but there are  much fewer links outside of it. 
These facts, which are crucial consequences of the  non-symmetric nature of interactions and the heterogeneities of the local connectivities,
are responsible for the non standard response of the system to noise. To see this, let us consider the extreme situation where  the network  consists exactly  of a small cluster of $n_c$ nodes {\it all} connected with one another, and a large homogeneous cluster of $N-n_c$ nodes with a few links pointing to the small one. In this case, the small cluster evolves dynamically completely independently of the large one and therefore exhibits
global fluctuations of order (according to Eqs. (\ref{perp})(\ref{dsym})) $\langle \left (\delta M^\perp \right )^2 \rangle \sim t/n_c$. Thus, it will change its state on short scales $\tau \sim n_c$. The large  cluster would by itself fluctuate much less, but due to the connections to the small cluster it is quickly dragged from its original direction and the relaxation time of the whole network is drastically decreased.

The occurrence of almost disconnected clusters has statistical origins. When we draw nodes at random in space and build the interaction graph, we can by mere chance produce such regions. The probability that one such region is formed is related to the probability of producing a clump of $n_c$ close nodes, i.e. a local heterogeneity.  This probability can be very small (depending - for example - on the value of $n_c$), but it only depends on the local properties of the network. Thus, as in typical nucleation processes, the larger the system the larger is the chance that somewhere in the network  one of such clumps occurs, explaining the growth of  the secondary peak with size. Close to the boundary nodes have neighbors only in half of  the available space, potentially increasing local clustering. 
Indeed we can quantitatively verify that the regions with large $u^0_i$ and clustering coefficient tend to be located at the periphery of the network, with typical normalized distance $r$ from the network center of order $r\sim 0.9$ 
($r=0$ corresponding to the center, and $r=1$ to the border, see Fig. \ref{fig:examples}b)).

So far we analyzed the ensemble of NERH.
Our results, however, rely on a few very general properties: an interaction graph, which is direct and with finite connectivity (local asymmetric interactions); an imitative dynamics (mutual alignment); the presence of a boundary. Besides, they are robust and qualitatively hold when changing the value of $n_c$ 
(see Fig.\ref{fig:othernc}).  For this reason we expect our results to hold for many real instances of coordinated behavior, providing a clear explanation of the mechanism leading to collective swings. In many biological cases, the interaction network is not fixed but evolves dynamically. Flocks of birds, for example, at a given instant of time have an interaction graph very similar to a NERH \cite{ballerini+al_08,bialek+al_12}. As birds move in space and exchange positions the graph will progressively change. A single flock explores  during its motion many different realizations of asymmetric random graphs, the dynamics therefore playing the role of the ensemble for the NERH. Whenever a `quick' graph with large diffusion coefficient is visited we expect the flock to spontaneously  change direction of motion. Asymmetric interactions can also modify the hydrodynamic behavior of very large flocks, as discussed in \cite{sriram} for the case of longitudinal asymmetries. Even though the focus of our paper is different, 
both works point out that non-symmetric interactions can have a fundamental role - distinct from motility - in the non equilibrium behavior of active systems.

Testing directly our predictions on real data - as we did for the NERH -  is not straightforward, because we do not know a priori the interaction graph between individuals.  In some cases, inference techniques allow to extract some average properties (e.g. the connectivity $n_c$) \cite{bialek+al_12,cavagna+al_15}, but retrieving the entire graph requires an experimental statistics which is not available to date. There are however  several experimental observations, which are consistent with our findings and support our explanation. 
Flocks of birds indeed exhibit spontaneous coherent  turns very frequently even for large group sizes \cite{attanasi+al_14}. Besides, all turns start from the lateral periphery of the flock and initiators are individuals displaying unusual and systematic directional fluctuations \cite{attanasi+al_15}, exactly as predicted by the NERH analysis. Recent results on fish schools \cite{rosenthal+al_15} show that these groups occasionally display spontaneous evasion waves. Also in this case initiators of the startle events are located peripherally and have a large clustering coefficient, in line with our results.

For a biological group, controlling and regulating collective behavior has a crucial role. The group must maintain a large sensitivity to perturbations to ensure efficient collective responses (as in anti-predatory maneuvers), and at the same time retain group coherence. The mechanism we described shows how to achieve such a marginal stability: the system is always highly ordered but off-equilibrium effects allow for rapid collective swings. 

\vskip 1 cm

\appendix

\begin{center}
\bf SUPPLEMENTARY MATERIAL
\end{center}

\section{Dynamical Equations}
Let us consider a generic interaction graph specified by the connectivity matrix $n_{ij}$, where $n_{ij}=1$ if node $i$ interacts with node  $j$ and $n_{ij}=0$ otherwise. For a Heisenberg model on a regular lattice, $n_{ij}=n_{ji}\ne 0$ only when $i$ and $j$ are neighboring nodes on the lattice. The equilibrium properties of the Heisenberg model on a lattice are fully specified by the Hamiltonian: 
\be
{\cal H}=- J \sum_{ij} n_{ij} {\vec \sigma}_i \cdot {\vec \sigma}_j \ .
\label{Heisenberg2}
\ee

There exist several possible dynamics that can be associated to this Hamiltonian, asymptotically leading to the same equilibrium behavior \cite{HH_77}. Here, we consider a general Langevin dynamics that has been widely studied in the literature of ferromagnetic systems, and has been found to quantitatively describe several kinds of active living systems and biological groups \cite{vicsek_review, marchetti_review,attanasi+al_14}. To simplify the notation, we will consider the case of planar spins, where the nodes of the network (the lattice sites in the case of a regular lattice) live in a $3$d space, but the vectors $\vec\sigma$ can only fluctuate in $2$ dimensions. The generalization to three dimensions is conceptually straightforward even though algebraically heavier (see, e.g. \cite{chaikin_00,cavagna+al_15}).

For planar spins we can represent each spin $\vec\sigma$ using the angle $\varphi_i$ formed by the spin with a given reference direction. In the ordered phase we conveniently choose this direction as the direction of the global order parameter $\vec{M}$.  Thus
\bea
\sigma_{i,x}&=&\cos(\varphi_i)\\
\sigma_{i,y}&=&\sin(\varphi_i)\\
\eea
The Hamiltonian can be easily rewritten in terms of the phases $\{\varphi_i\}$. In particular, at low temperature the system is highly ordered, the phases are small and expanding we get
\be
{\cal H}
\sim \frac{J}{2} \sum_{ij}n_{ij}(\varphi_i-\varphi_j)^2
\label{Heisenberg3}
\ee

The dynamical behaviors we will consider, can  be derived from the Langevin dynamics associated to this Hamiltonian. When considering a full Langevin dynamics, including second order inertial terms and first order dissipative terms, we have
\be
\chi \frac{d^2 \varphi_i}{d t^2}=-J\sum_j \Lambda_{ij}\varphi_j-\eta\frac{d\varphi_i}{dt}+\xi_i
\label{eq:dynsup}
\ee
where $\Lambda_{ij}=-n_{ij}+\delta_{ij} \sum_{k} n_{ik}$ is the discrete Laplacian. The coefficients $\chi$ and  $\eta$ represent, respectively, a rotational inertia and a rotational viscosity; and $\xi_i$ is a random Gaussian noise  
with $\langle \xi_i \xi_j\rangle = \delta_{ij} \Delta \  \eta$.  For a symmetric interaction matrix, and for the Heisenberg model on a regular lattice in particular,  this dynamics approaches at large times  the stationary Boltzmann distribution determined by Hamiltonian (\ref{Heisenberg3}), and temperature $T=\Delta/(2)$.   The dynamics described by Eq. (\ref{eq:dynsup}) can however be defined more generally even when $n_{ij}$ is not symmetric, independently of the existence of an equilibrium distribution. This is  what usually happens in many biological systems, where interactions are not reciprocal and individuals move updating their behavioral variables subject to an imitative/alignment force exerted by neighbors, as described in Eq.~(\ref{eq:dynsup}).   For this reason, we shall consider from now on the general case where $n_{ij}$ and $\Lambda_{ij}$ are 
not necessarily symmetric matrices.

By taking the limit $\chi/\eta^2 \to 0$  we recover a purely over-damped dynamics, which has been widely considered in the literature of magnetic systems. It has been used for example in \cite{goldschmidt_86,niel_86} to investigate the critical dynamics and finite size scaling in the Heisenberg ferromagnet, and it corresponds to  model A in the renowned dynamical classification of  \cite{HH_77}.  The same kind of over-damped equation is also adopted in many models of active systems where it  describes the dynamical update rule for the velocities of self-propelled interacting particles (see e.g. the Vicsek model and its variants \cite{vicsek_95,tu+al_98,couzin+al_02,chate+al_04,chate+ginelli_10,vicsek_review} . 
The limit $\eta\to 0$ corresponds instead to a purely conservative dynamics,  as in model F of \cite{HH_77}  and - for  3d spins - to model G \cite{HH_77} and the rotor spin model  \cite{chaikin_00}). Natural flocks of birds have also been shown to be appropriately described by Eq. (\ref{eq:dynsup}) in the deeply underdamped regime (small $\eta^2/\chi$) \cite{attanasi+al_14,attanasi+al_15, cavagna+al_15}.

To solve the dynamical equations it is convenient to rewrite Eqs.~(\ref{eq:dynsup}) as a set of first order equations. Let us introduce a new variable $s_i=\chi \ (d\varphi_i/dt)$. In terms of $\{\varphi_i;s_i\}$ Eq. (\ref{eq:dynsup}) becomes \cite{cavagna+al_15}
\bea
\frac{d\varphi_i}{dt}&=&\frac{1}{\chi}s_i  \label{eq:uno}\\
\frac{ds_i}{dt}&=&-J \sum_j \Lambda_{ij}\varphi_j-\frac{\eta}{\chi} s_i+\xi_i \label{eq:due}\ .
\eea

In the $\eta\to 0$ limit, when dissipation and noise disappears, these equations clearly show the conservative nature of the underdamped dynamics.The variable $s_i$ satisfies in this case a continuity equation  (Eq.~(\ref{eq:due})), and the global variable $S=1/N\sum_i s_i$ is conserved given that, by construction, the discrete Laplacian satisfies the constraint $\sum_j \Lambda_{ij}=0$. This conservation law can also be seen as a consequence of the Hamiltonian structure of the equations when $\eta=0$.  We can interpret $s_i$ as the conjugate momentum to the phase $\varphi_i$ (i.e. the generator of the rotational invariance of the spins) and add a kinetic term $\sum_i s_i^2/(2\chi)$ to the Hamiltonian. Then, Eqs. ~(\ref{eq:uno})(\ref{eq:due}) are nothing else than the Hamilton equations for $\varphi_i$ and $s_i$ \cite{cavagna+al_15}

We can recast Eqs.~(\ref{eq:uno})(\ref{eq:due}) in vectorial notation
\be
\frac{d\Psi}{dt} =- L \Psi + \Xi
\ee
where
\be
\Psi=\left ( \begin{array}{c} 
\varphi \\ s \end{array} \right ) 
\quad
\Xi=\left ( \begin{array}{c} 
0\\ \xi \end{array} \right ) 
\quad
L=\left ( \begin{array}{cc} 
0 & -1/\chi\ \mathbb{I}\\
J \Lambda & \eta/\chi\ \mathbb{I} 
\end{array} \right ) 
\ee
and $\varphi=(\varphi_1\cdots\varphi_N)$, $s=(s_1\cdots s_N)$, $\xi=(\xi_1\cdots \xi_N)$ are $N$-dimensional vectors, and $\mathbb{I} $ is the identity matrix in the $N$-dimensional space spanned by the vector $s$.

The formal solution of this linear equation is easily obtained by standard methods \cite{bender}, giving
\bea
&&\Psi(t)=\exp{(-tL)}\Psi_0+\int_0^t dt' \exp{\{-(t-t')L\}}\ \Xi(t') \nonumber\\
\nonumber \\
&&\Psi_0= \Psi(t=0)
\eea

This solution can be expressed in terms of the eigenvectors and eigenvalues of the matrix $L$, and consequently - of the eigenvectors and eigenvalues of $\Lambda$. Let us assume for the sake of generality that $\Lambda$ is not symmetric (we will look specifically at the symmetric case below), and let us call $\{\lambda\}$ its eigenvalues, and  $\{v^\lambda\}$ and $\{u^\lambda\}$  the corresponding right and left $N$-dimensional eigenvectors.  Then,  after some algebra, we get
\bea
&&\varphi_(t)=v_i^0 \int_0^t dt' \ \frac{1-e^{-2\gamma (t-t')}}{\eta}  \sum_k u^0_k \xi_k(t') \nonumber \\
&& + \sum_{\lambda > 0} \sum_k v_i^\lambda u_k^\lambda \int_0^t dt' e^{-\gamma(t-t')} \ \frac{\sin\left [ \omega(\lambda)(t-t')\right ]}{\chi \omega(\lambda)}\xi_k(t') \ .
\nonumber\\ 
\label{eq:sol}
\eea
with
\be
\gamma=\frac{\eta}{2\chi} \quad\quad c_s^2=\frac{J}{\chi} \quad \quad  \omega(\lambda)=\sqrt{\lambda c_s^2-\gamma^2}
\ee

In Eq.~(\ref{eq:sol}) we have explicitly separated the contribution of the zero mode $\lambda=0$, from the contributions of the non-zero eigenvalues. We remind that the Laplacian matrix $\Lambda$ has a zero mode due to the diagonal constrain. Indeed $\sum_j \Lambda_{ij}=0$ by construction, which implies that $\lambda=0$ is an eigenvalue and $v^0_i=const$ the corresponding right eigenvector. This zero mode plays a crucial role, and has a deep physical meaning: it derives from the original invariance of the Hamiltonian and of the dynamical equation with respect to rotations of the spins $\{\vec{\sigma}_i\}$, which in turn implies the translational invariance in terms of the phases $\{\varphi_i\}$. This mode is usually referred to as the Goldstone mode in the literature on the Heisenberg model, and describes a marginal manifold of soft `easy' fluctuations. 

Let us now investigate the dynamical behavior of the global magnetization. Let us assume that our system is initially equilibrated in a highly ordered phase with magnetization $\vec{M}(t=0)=M_0 \vec{n}$. Due to noise, the magnetization fluctuates and $\vec{M}(t)$ drifts away from its original direction.
Let us now call $\varphi_i$ the angle formed by the spin $\vec{\sigma_i}$ with $\vec{n}$. Then, in the regime where the phases are small enough to expand we have 
\bea
 \vec{M}(t)&=&(\vec{M}\cdot\vec{n})\vec{n}+\vec{\delta M}^\perp \label{eq:mperp}\\
(\vec{M}\cdot\vec{n}) &\sim &\frac{1}{N}\sum_i (1-\varphi_i(t)^2) \label{eq:exp1}\\
\delta M^\perp &\sim&\frac{1}{N} \sum_i \varphi_i \label{eq:exp2}
\eea
The perpendicular component of $\vec{M}(t)$  describes how much the magnetization is far from the initial direction $\vec{n}$. When $\delta M^\perp$ becomes of order one, we can say that the system has relaxed to a different state from the original one. The relaxation time $\tau$ can be therefore defined as the time when $(\delta M^\perp)^2\sim O(1)$. Let us now compute $(M^\perp)^2$ from the dynamical solution (\ref{eq:sol}) and compute the relaxation time.
\bea
&& (\delta M^\perp)^2 = \frac{1}{N^2}\sum_{ij}\langle \varphi_i\varphi_j \rangle =
\left ( \frac{1}{N} \sum_i v^0_i \right )^2  \sum_j (u^0_j)^2 \  \ \frac{\Delta t}{\eta} \nonumber \\
&&\quad +  \sum_{\lambda >0} \left ( \frac{1}{N} \sum_i v^\lambda_i \right )^2 \sum_j (u^\lambda_j)^2 \Delta  \left \{ \frac{1}{J\lambda} 
 + \frac{e^{-\frac{\chi}{\eta} t}\eta}{\chi^2\omega(\lambda)^2} \right. \nonumber \\
&& \quad\quad \left.  \times  \left [ -\frac{\chi}{2\eta}+\frac{  \frac{\eta}{2\chi\omega{\lambda}}\cos[2\omega(\lambda) t]-\frac{1}
{\omega(\lambda)}\sin[2\omega(\lambda)t]}{4+[\eta/(\chi\omega(\lambda)]^2} \right ]\right \} \nonumber \\
&& \quad\quad + \cdots
\label{eq:mperp2}
\eea
where we have discarded the terms related to the initial conditions and the off-diagonal terms involving double sums over different eigenmodes, which are all sub-leading terms in time.

Let us now consider separately the two cases of symmetric and asymmetric interactions, which - as we shall see  - might give rather different results.

\subsection{Symmetric Interactions}

If the matrices $n_{ij}$ and $\Lambda_{ij}$ are symmetric, their left and right eigenvectors coincide. We therefore have $u^\lambda_i=v^\lambda_i$ for all $i=1\cdots N$. In particular, this implies that
\be
u^0_i=v^0_i=w^0_i=\frac{1}{\sqrt{N}}
\ee
and, by the orthogonality condition between different eigenvectors,
\be
\sum_i v^\lambda_i=0  \quad\quad \lambda\neq 0
\ee  

In this case, therefore, all terms in Eq.~(\ref{eq:mperp}) go to zero, but the contribution of the zero mode. Thus we get
\be
(\delta M^\perp)^2 = \frac{\Delta}{N}  \frac{t}{\eta},
\ee
which shows that the magnetization departs from the original direction with a diffusive behavior along the zero eigen-space. To stigmatize such behavior we can introduce a diffusion coefficient $D_0$ and write
\be
(\delta M^\perp)^2 = D_0 \frac{t}{\eta}, \quad\quad\quad D_0=\frac{\Delta}{N}
\ee
This immediately gives the expression of the relaxation time for a system of size $N$
\be
\frac{\tau_N}{\eta} \sim \frac{1}{D_0}= N
\ee

Thus, independently of the specific kind of dynamics followed by the system, when interactions are symmetric the relaxation time grows linearly with system size. We note that the only property necessary to obtain this result is the equivalence of the left and right zero eigenvectors, stemming from the symmetry condition. Thus, this result not only holds for the regular lattice case, but also for any kind of symmetric interaction graph (as long as there is only one connected component) \cite{cassi}. So far, we computed the relaxation time in the deeply ordered regime, where the expansion with respect to the phases $\{\varphi_i\}$ can be performed. Close to the critical region the computation gets more complex, involving renormalization group techniques. It has been performed for the regular lattice case in \cite{goldschmidt_86,niel_86} leading to the same scaling of relaxation time with size.

\subsection{Non-symmetric interactions}

When interactions are not symmetric, the right and left eigenvectors are not the same. When we consider the zero eigenmode, we still have that the right eigenvector must be a constant (precisely because $\sum_j \Lambda_{ij}=0$ by construction), but the same is not true for the left eigenvector.
We thus have
\be
v^0_i=const   \quad\quad u^0_i\neq v^0_i \quad\quad \sum_i v^0_i u^0_i =1 
\ee
and
\be
\sum_i u^\lambda_i=0 \quad\quad \sum_i v^\lambda_i\neq 0  \quad\quad \lambda\neq 0
\ee
As a consequence, the contribution of the zero mode does not necessarily scale as $1/N$ and all the terms in the r.h.s. of Eq.~(\ref{eq:mperp}) are in principle different from zero. For large times, however, the dominant term is always the first one (the only one growing with time), and we get (setting $v^0_i=1$)
\be
(\delta M^\perp)^2 = D \frac{t}{\eta}, \quad\quad\quad D=\frac{\Delta}{N}\sum_i (u_i^0)^2
\ee
and
\be
\frac{\tau}{\eta}\sim \frac{1}{D}
\ee
Now the scaling of the relaxation time with size depends on the localization properties of the left eigenvector $u^0$, and - therefore - on the topological properties of the interaction graph.

\section{Statistical Analysis of the  Non-symmetric Euclidean Random Heisenberg Ensemble}

 \begin{figure}[t]
\includegraphics[width=0.65\columnwidth]{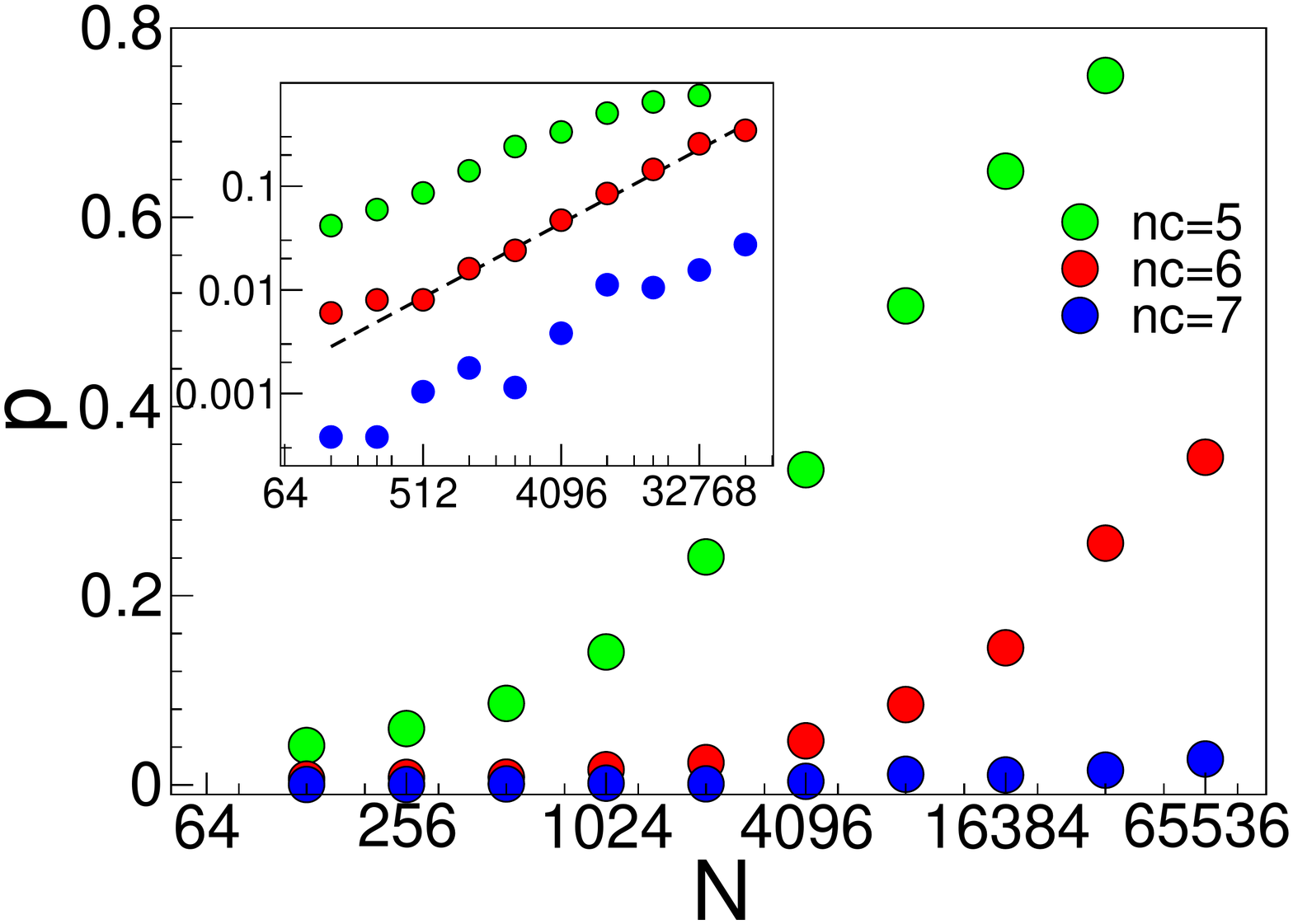}
\includegraphics[width=0.7\columnwidth]{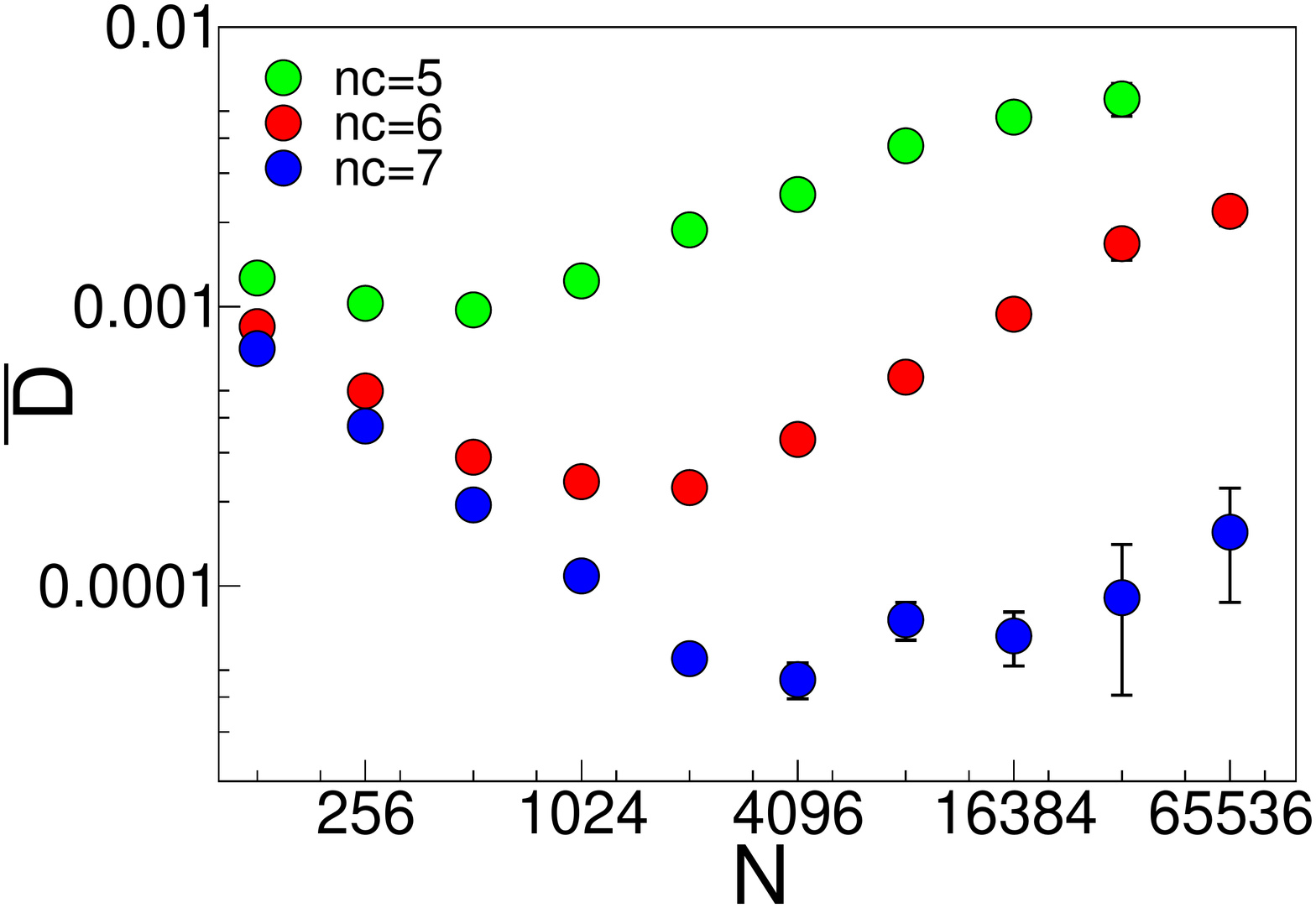}
\caption{a)  Probability of the secondary peak as a function the network size. Inset: same plot in log-log scale; the dotted line corresponds to a linear fit, indicating that $p$ initially increases as a power law with exponent $ 0.79 \pm 0.02$. c)  Average diffusion coefficient as a function of the network size. $N\in [128: 65536]$; $n_c=5,6,7$.}
\label{fig:othernc}
\end{figure}

To investigate the properties of the Non-symmetric Euclidean Random Heisenberg Ensemble (NERH), we generate Euclidean Random Matrices of given size $N$ and connectivity $n_c$ in the following way. We draw $N$ points uniformly in a three-dimensional sphere. Then, given a point/node $i$ we identify its first $n_c$ neighbors and establish a directed link (corresponding to $n_{ij}=1$) between node $i$ and each of such neighbors. We repeat the procedure for all the nodes. Nodes which are not connected by any link have $n_{ij}=0$. 
In a biological context, for example in a group of animals, we would say that $n_{ij}=1$ when individual $i$ sees/perceives/tracks neighbor $j$, and $n_c$ therefore represents the global number of neighbors $i$ is taking care of. In network terminology,  $n_c$ is called the in-degree of node $i$, i.e. the number of other nodes  by which $i$ is influenced.  

\subsection{Laplacian diagonalization}

To compute all the eigenvectors and the eigenvalues of the Laplacian we should in principle diagonalize the matrix $\Lambda$ (or, equivalently, the matrix $n_{ij}$). However, diagonalization can be numerically demanding, especially when the size of the system is large. Fortunately, if we are interested in computing the diffusion coefficient $D$ we only need to compute the zero left eigenvector. This turns out to be much simpler, because we can use power iteration: starting from a uniform vector and iteratively applying the transition matrix $P_{ij}=n_{ij}/n_c$ the process will converge to the zero left eigenmode.
To see this, we note that $P$ has the same eigenvectors of $\Lambda$, and eigenvalues $\mu=1-\lambda / n_c \leq 1$. The zero mode  of the Laplacian therefore corresponds to the maximal eigenvalue of $P$, i.e. $\mu=1$. Let us now consider a vector $x$ such that $\sum_i v^0_i x_i = \sum_i x_i =1$.  We then have
\be
\lim_{n\to\infty} x P^n \to  u
\ee
since the projection of $x$ on any mode with $\mu<1$ goes to zero as $n$ increases. This allows to implement an easy iterative procedure. We choose a uniform vector $x$ with $x_i=1/N$ for all $i=1\cdots N$. Then, we apply the transition matrix to get $x'=xP$ and we continue to iterate until the difference $|x'-xP|$ is sufficiently small. The threshold for convergence was set when the difference between two successive iterations was smaller then $10^{-9}$. We used 
a Kahan algorithm to improve numerical precision. In this way, we were able to compute the zero left eigenvector for sizes of up to $N=65536$. We verified the accuracy of the method by comparing the results given by power iteration with the exact diagonalization for sizes of up to $N=4096$.

\subsection{Analysis at different values of $n_c$}

In the main text we presented and discussed results obtained for  NERH ensembles at different values of $N$ and for $n_c=6$. Qualitatively analogous results also hold for different values of $n_c$. In Fig.~\ref{fig:othernc} we report the behavior of the probability of the secondary peak, and of the average diffusion coefficient (same plot as in Fig.2), for $n_c=5,6,7$. Also for $n_c=5$ and $n_c=7$  the peak weight and the average diffusion coefficient increase with the system size. A large value of $n_c$ makes the whole network more connected, and - for a given size $N$ - we expect the occurrence of almost disconnected clusters to be statistically less probable than at smaller $n_c$. Indeed, we can see that for $n_c=7$ much larger sizes, as compared to $n_c=5,6$  are needed to make the secondary peak gain weight. Still, the behavior is qualitatively the same, as well illustrated in the log-log plot in the inset: the probabilities exhibit different amplitudes but they increase with the same exponent.

\subsection{Clustering properties and spatial distribution of influent nodes}

\begin{figure}[t!]
\includegraphics[width=0.49\columnwidth]{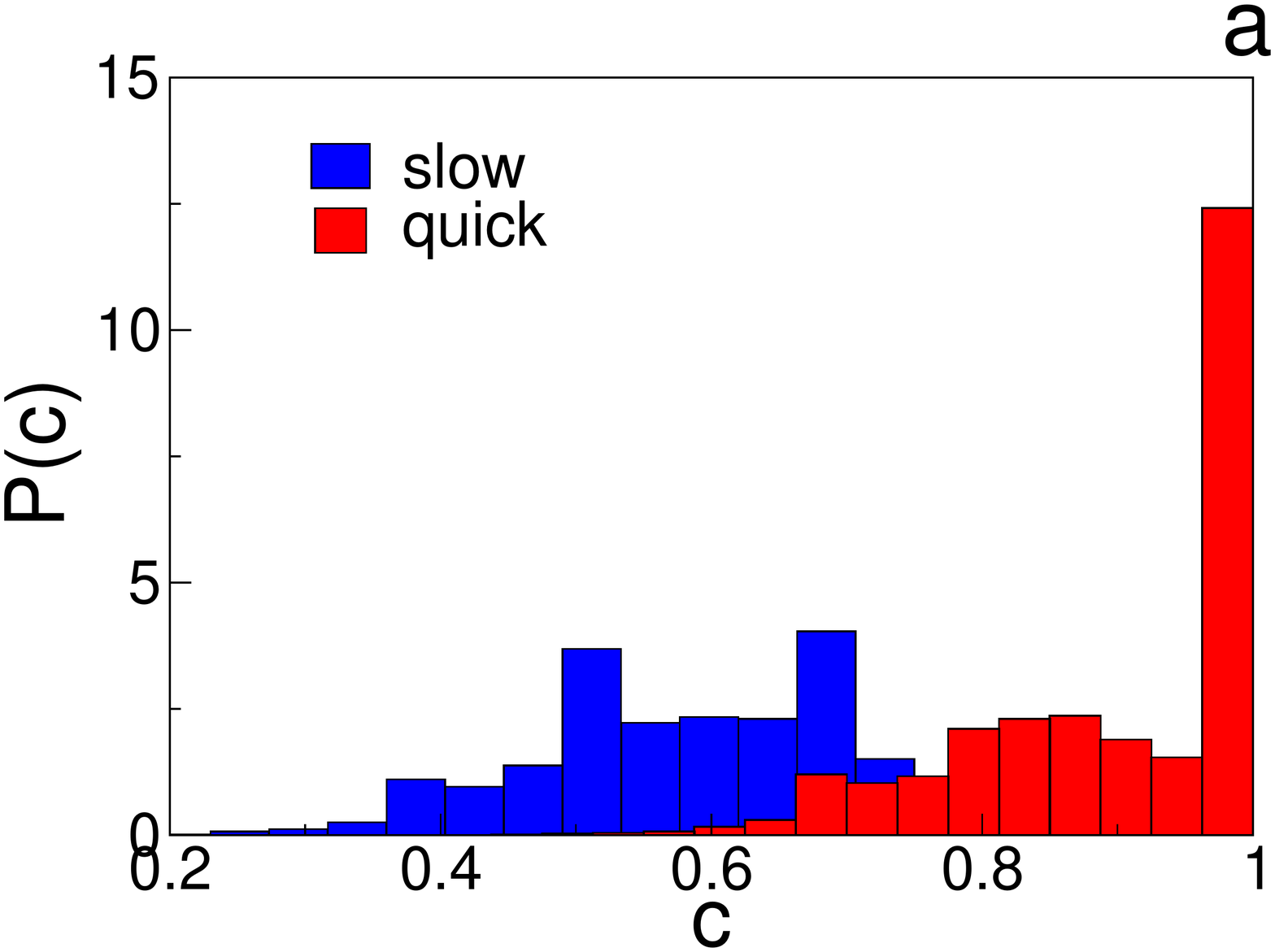}
\includegraphics[width=0.49\columnwidth]{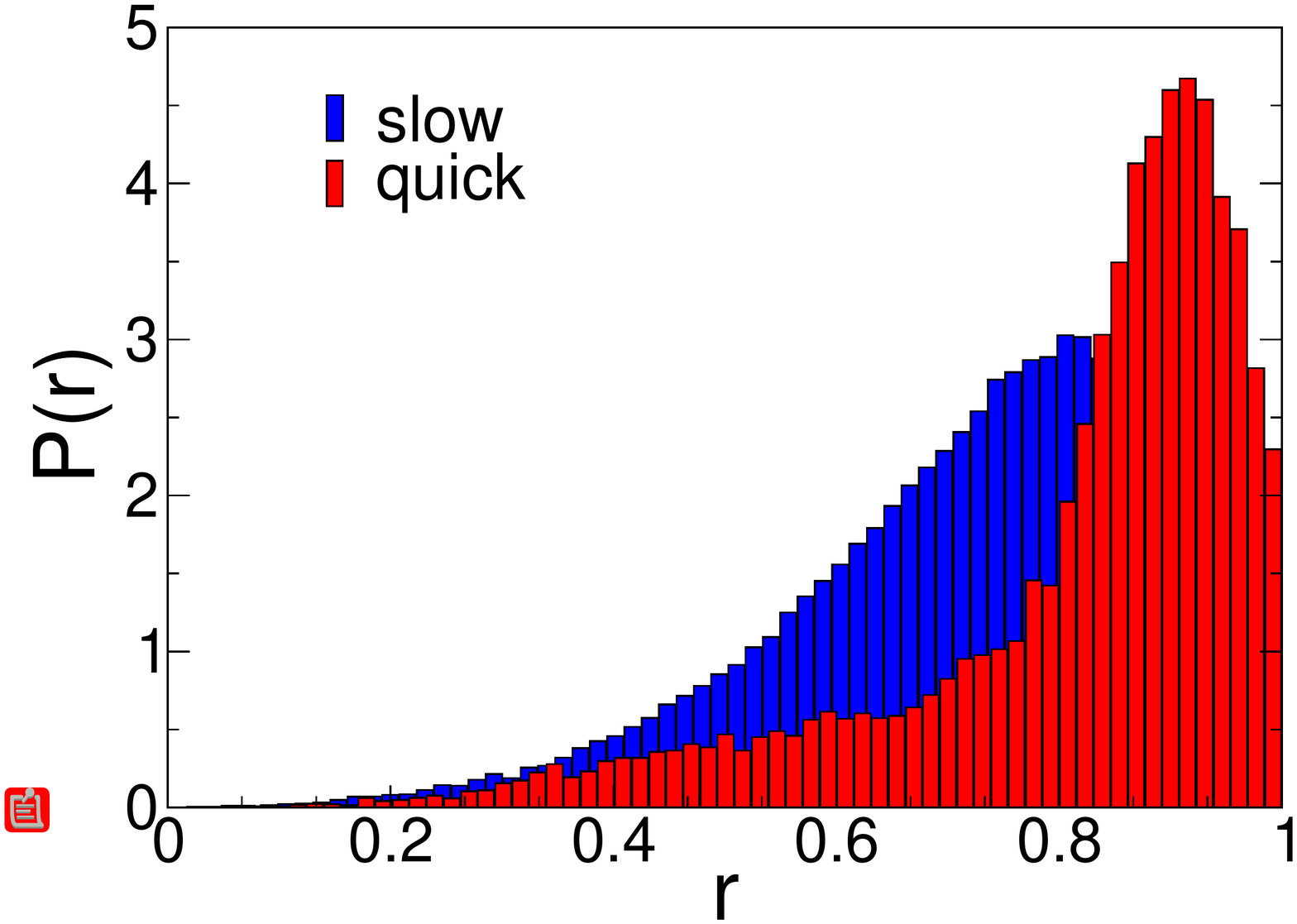}
\caption{a) Distribution probability of the clustering coefficient for the nodes mostly contributing to the zero eigenmode (the first $n_c$ nodes with highest centrality), for the quick networks (red) in the $N=1000$ ensemble. We also report the corresponding distribution for the slow networks (blue) for comparison.  b) Distribution probability of the normalized distance from the network's center for the same nodes ($r=0$ corresponds to the center, $r=1$ to the border).}
\label{fig:examples}
\end{figure}

One interesting question is to understand where the most influent nodes, i.e. the nodes that mostly contribute to the zero-mode fluctuations, are located in the network and one with respect to the other. One informative quantity in this respect is the so-called clustering coefficient of a given node $i$, defined as
\be
c_i=\frac{1}{n_c(n_c-1)}\sum_{jk} n_{ij}n_{ik}n_{jk} \ .
\ee
This coefficient is large, and close to $1$, if node $i$ and its interacting neighbors share their respective interacting neighbors, i.e. it is an indication of how much clustered the network is around node $i$. The distribution probability of the $c_i$ for the most influent nodes (the first $n_c$ nodes with largest weight $(u^0_i)^2)$ is plotted in Fig.\ref{fig:examples}, for both quick networks (large $D$) and slow networks (small $D$). To compute this distribution, we first computed the zero left eigenmode $\bf u^0$, and the associate partecipation ratio $PR=1/N \sum_i (u^0_i)^2$. For quick networks $PR$ is of order $1/n_c$ indicating that the zero eigenmode is localized approximately on $n_c$ nodes. Thus, we rank the nodes according to the value of their weight $(u^0_i)^2$ and select the first $n_c$ nodes. For each of such nodes we computed the clustering coefficient $c_i$, and, finally, its distribution. We adopted the same procedure also for slow networks for consistency. Fig.\ref{fig:examples}a) shows that for quick networks the most influent nodes tend to have very large values of $c_i$. We conclude that these nodes form a clustered region, and are very interconnected one to the other. 
 
To understand where these $n_c$ most influent nodes are located in the network, we can look at their position with respect to the network center. In order to do so, for each one of such nodes we compute  the normalized distance $r$ with respect to the network center. To build NERH networks - we remind - we initially draw points uniformly in a spherical region in three dimensions. We can therefore define the normalized distance $r$ as the distance of the node from the center, divided by the maximal possible distance (i.e. the radius of the spherical region where all nodes are located). In this way, $r=1$ corresponds to a node located exactly at the outer edge of the network, $r=0$ to a node located in the center. The distribution of $r$ is plotted in Fig.\ref{fig:examples}b), and shows that the most influent nodes are most of the time located close to the boundary of the network.

\section{Numerical simulations}

To validate the relationship between diffusion coefficient and relaxation time, we implemented the dynamical equations described in the main text using numerical simulations. We considered  the dynamical model described by Eq. (\ref{eq:dynsup}) in the more general case, where the spins are vectors fluctuating in the whole three-dimensional space. In three dimensions, there are three phases describing rotations of the spin $\vec{\sigma}$ around each cartesian axis, and three associated auxiliary momentum variables $s_i$, that can be conveniently represented by a $3d$ vector $\vec{s}_i$.  

 The generalization of Eqs.~(\ref{eq:uno})~(\ref{eq:due}) - expressed in terms of the spins $\{ \vec{\sigma_i}\}$ and the momenta $\{\vec{s}_i\}$ then read \cite{cavagna+al_15}
\bea
\vec{\sigma}_i(t+dt)&=& \vec{\sigma}_i(t)+ \frac{1}{\chi} \vec{s}_i(t) \times  \vec{\sigma}_i(t) \, dt \label{eq:num1}\\
\vec{s}_i(t+dt) &=& \vec{s}_i(t)+ \vec{\sigma}_i(t) \times J \sum_{j} n_{ij} \vec{\sigma}_j(t) \, dt \nonumber \\
&-&\frac{\eta}{\chi}s_i(t) \, dt +  \vec{\sigma}_i(t) \times \vec{\xi}_i(t) \, \sqrt{dt}
\label{eq:num2}
\eea
where we used a discrete time update formulation, which can be implemented numerically. In Eq.(~\ref{eq:num2}) the noise is a full three-dimensional Gaussian noise, with variance $\langle \vec{\xi}_i(t)\cdot\vec{\xi}_j(t^\prime)\rangle = \Delta \delta_{ij}\delta(t-t^\prime) \eta$. In this case, we define the temperature as $T=\Delta / (2 d)$.

 These equations correspond to a lattice version of the Inertial Spin Model introduced and discussed in \cite{cavagna+al_15} to describe flocks of birds. This dynamical model has been investigated in \cite{cavagna+al_15}, where a systematic analysis of the different regimes (underdamped vs overdamped) has been performed for a self-propelled particle model. 
 
 To investigate the dynamical behavior of the NERH networks discussed in the main text, we chose the values of the different parameters in such a way as to keep the system always in a large magnetization regime. Several values of $\eta$ and $\chi$ were considered, corresponding to underdamped and overdamped dynamics. The exact values of the parameters are reported in the main text. For each network, the system was initialized with polarization  one and the dynamical evolution implemented through Eqs.~(\ref{eq:num1})(\ref{eq:num2}). After a transient of the order of the relaxation time of the orientational correlations, we fixed a reference initial time $t_0=0$ and evaluated the magnetization $\vec{M}_0$ and its direction $\vec{n}$. Then we computed $\vec{M}(t)$ and its perpendicular component $\delta \vec{M}^\perp (t)$ as defined in Eq.(\ref{eq:mperp}). Finally, we computed $(\delta M^\perp(t))^2$ and averaged over all possible initial times $t_0$ along the simulation. Typical results are plotted in Fig.1a for three different networks. To compute the relaxation time $\tau$, we used a reference threshold and determined when the averaged 
 $(\delta M^\perp(t))^2$  reached the threshold. The results plotted in Fig.1b are obtained for a threshold of $0.3$; qualitatively similar results are obtained also for different values of the threshold. 
Since the relaxation time is proportional to $\eta$, simulations in the overdamped regime are much slower and numerically more demanding. For the same reason, given our maximum simulation time, the dynamical averages of $(\delta M^\perp(t))^2$  for a single network (and, consequently, the relaxation time) tend to be more fluctuating. Thus, to improve the statistics, when comparing relaxation time and diffusion coefficient, we also averaged the relaxation time over different networks with similar diffusion coefficient. Results are reported in Fig.1b, error bars in the figure correspond to fluctuations over networks in the same $D$ bin.

%
{\bf Acknowledgments.}
 We thank V. Martin Major, T. Mora, S. Ramaswamy and A. Walczak for discussions, and S. Ramaswamy for sharing his preliminary draft of Ref. \cite{sriram}. This work was supported by grants IIT--Seed Artswarm, ERC--StG n.257126 and US-AFOSR - FA95501010250 (through the University of Maryland).





\begin{thebibliography}{99}


\bibitem{attanasi+al_14} Attanasi A. et al,
{\it Nat. Phys.} {\bf 10}, 691--696 (2014) 

\bibitem{attanasi+al_15}
Attanasi A. et al,
{\it J. R. Soc. Int} {\bf 12}, 20150319 (2015)

\bibitem{centola_10} Centola D., 
 {\it Science} {\bf 329} 1194--1197 (2010).

\bibitem{couzin+krause_03}
Couzin, I.~D. \& Krause, J.,
 \textit{Adv.  Study  Behav.} {\bf 32}, 1--75  (2003).

\bibitem{vicsek_review} 
Vicsek T, Zafeiris A, 
{\it Phys. Rep.} {\bf 517}, 71-140 (2012)


\bibitem{bialek+al_12}
W.~Bialek {\em et~al.}, 
{\it Proc. Natl. Acad. Sci.} {\bf 109}, 4786 (2012); {\it Proc. Natl. Acad. Sci.} {\bf 111}, 7212 (2014).



\bibitem{goldschmidt_86} Goldschmidt YY, 
 {\it Nucl. Phys. B} {\bf 280} 340-354 (1987)

\bibitem{niel_86}Niel, J.C. \& Zinn-Justin, J., 
{\it Nuclear Physics B } {\bf 280} 355-384 (1987)

\bibitem{HH_77} Hohenberg, P.C. \& Halperin B.I.,
{\it Rev.  Mod. Phys.}  {\bf 49}  435 (1977).

\bibitem{chaikin_00} Chaikin, P M., Lubensky TC. Principles of condensed matter physics. Vol. 1. Cambridge University Press, Cambridge (2000).


\bibitem{vicsek_95}Vicsek T. et al. 
 {\it Phys. Rev. Lett.} {\bf 75} 1226 (1995)


\bibitem{cavagna+al_08}
Cavagna A. {\it et al.}, 
{\it Math. Biosc.} {\bf 214}, 32--37 (2008)


\bibitem{ballerini+al_08}
Ballerini M. et al.
  \textit{Proc. Natl. Acad. Sci. USA} \textbf{105}, 1232--1237 (2008).
  


\bibitem{cavagna+al_15}
Cavagna A. et al.
{\it Phys. Rev. E} {\bf  92} 012705 (2015)

\bibitem{mora+al_15} 
Mora, T et al.
arXiv preprint arXiv:1511.01958 (2015).



\bibitem{cavagna+al_14} Cavagna A, et al.
 {\it J. Stat. Phys.},  DOI: 10.1007/s10955-014-1119-3 (2014).


\bibitem{cavagna+al_15b} Cavagna A, et al.
{\it Phys. Rev. Lett.} {\bf 114}, 218101 (2015)


\bibitem{sriram} Dadhichi L.P., Chaijwa R., Maitra A.\& Ramaswamy S., preprint (2016).


\bibitem{zwanzig}
Zwanzig, R. 
Nonequilibrium statistical mechanics, 
Oxford University Press, Oxford (2001)

\bibitem{naomi}Young G, Scardovi L, Leonard N 
{\it Proc Am Control Conf} ACC, 6312Ð6317 (2010)


\bibitem{mezard} M\'ezard M., Parisi G., Zee A., 
 {\it Nuclear Physics B} {\bf  559}, 689-701 (1999)

\bibitem{skipetrov_11} Goetschy A. \& Skipetrov S.E., 
{\it Phys. Rev. E} {\bf 84} , 011150 (2011)

\bibitem{bonacich_01} Bonacich P., Lloyd P., 
{\it Social Networks} {\bf 23}, 191Ð201 (2001).


\bibitem{cavagna_review} Cavagna A.,
{\it Phys. Rep.} {\bf 476}, 51--124  (2009).
    
 \bibitem{rosenthal+al_15}
 Rosenthal SB et al.,
 \textit{Proc. Natl. Acad. Sci. USA} \textbf{112}, 4690-4695 (2015).




\bibitem{marchetti_review} Marchetti MC, Joanny JF, Ramaswamy S, Liverpool TP, Prost J et al. 
Hydrodynamics of soft active matter {\it Rev. Mod. Phys.} {\bf 85} 1143  (2013). 

\bibitem{tu+al_98}Y. Tu, J. Toner, M. Ulm
{\it Phys. Rev. Lett.}{\bf 80}, 4819 (1998)

\bibitem{couzin+al_02} I. D. Couzin et al., {\it J. Theor. Biol.}  {\bf 218}, 1 (2002).

\bibitem{chate+al_04}  Gr\'egoire, H. Chat/'e  {\it  Phys. Rev. Lett.}{\bf 92},
025702 (2004).

\bibitem{chate+ginelli_10} Ginelli F  \& Chate H. . {\it  Phys. Rev. Lett.}{\bf 105},
168103 (2010).

\bibitem{bender} Bender C.M., \& Orszag S.A. {\it Advanced mathematical methods for scientists and engineers I. }Springer Science \& Business Media, 1999.

\bibitem{cassi} Cassi D, 
{\it Phys. Rev Lett.}  {\bf 68} (1992)



\end{thebibliography}
\end{document}